\documentstyle[amssymb,aps]{revtex}

\begin{document}
\author{M. T. Hussein, R. Elmualed, N. M. Hassan}
\address{{Physics Department,Faculty of Education for Girls, Jeddah, K.S.A. }}
\title{{\normalsize Electron - Proton Scattering as a Probe of Nucleon Structure}}
\maketitle

\begin{abstract}
The problem of electron-proton scattering is handed over both the elastic
and inelastic scattering. Two models are presented in this sense. The first,
depends on the multi photon exchange ladder diagram, where the transition
matrix is expanded in multi steps form. The second model uses the multi
peripheral mechanism developed for the electromagnetic field. It allows the
particle production in the inelastic scattering processes. An iterative
procedure is found and inserted in a Monte Carlo program to reproduce the
differential cross section of the reaction. The comparison with the
experimental data shows bid fair in most cases.
\end{abstract}

\section{Introduction}

Recent experiments for electron-proton (ep) scattering, SLAC-E-136 [1],
DESY-HERA-HE [2], SLAC-E-133 [3] and others [4], played important role in
probing the nucleon structure and to reveal the dynamic mechanism of the
electron interactions inside the nucleon bag. Many trials have been done
before by a wide variety of empirical and theoretical models, ranging from
form factor scaling [5], and vector meson dominance [6] to quark- parton
models [7] and perturbative QCD [8]. An eikonal optical picture [9-10] was
used, based on multiple scattering of the incident electron with the
constituent valance quarks of the target hadron, assuming different forms of
the (electron-quark) binary wave functions.

In this article we shall deal with the problem of elastic and inelastic
collisions of the electron with proton from a different point of view. In
the following, we present two pictures for the (ep) scattering. The first is
a scattering in view of multi-photon exchange mechanism (MPEM) which is very
convenient for the elastic scattering in a wide range of momentum transfer
square $Q^{2}$. On the other hand, the multi-peripheral model (MPM) is
relevant to the inelastic and deep inelastic scattering and so works
properly for the problems of multi particle production. The article
structure goes as follows. In section 2, we present the postulates of the
MPEM, the formulation and the scenario of the MPM are given in section 3.
The Monte Carlo generators are summarized in section 4. Finally results and
discussion are given in section 5, followed by conclusive remarks.

\section{The multi photon exchange model (MPEM)}

It is assumed that elastic scattering of electrons on proton proceeds via
multi step processes represented by ladder diagrams [11]. We proceed on the
bases of the Feynman formalisms, assuming electromagnetic interaction acting
at each vertex of the ladder diagram Fig.(1). Expanding the transition
matrix $T$ of the (ep) scattering in terms of transition ladder diagrams $%
T^{(n)}$ so that,

\begin{equation}
T=\prod_{i}^{n-1}T_{i}
\end{equation}

$\{c^{n}\}$ and $\{T^{(n)}\}$ are the coefficients of expansion and
transition matrices of the adder diagram of order $n$. $T^{(n)}$ has the
form,

\begin{equation}
T^{(n)}=\int \cdot \cdot \cdot \int \prod_{1}^{n-1}dk_{j}\frac{%
V_{j+1,j}V_{j,j-1}}{k-k_{j}+i\epsilon }.
\end{equation}

Where the factor $\frac{1}{k-k_{j}+i\epsilon }$ stands for the Green's
propagator of the virtual intermediate state number $j$. $V_{j,j-1}$ is the
transition probability from the state $j-1$ to $j$, so that, $%
V_{j,j-1}=<\phi _{j}|V|\phi _{j-1}>.$ The screening Coulomb field working at
each vertex has the form, $V=-U_{0}\frac{e^{-\alpha r}}{r}$\ so that, each
internal integration, $V_{kk^{\prime }}$ has the form;

\begin{equation}
<k|V|k^{\prime }>=-U_{0}[2\pi ^{2}(\alpha ^{2}+|k-k^{\prime }|^{2})].
\end{equation}

Hence, a one step ladder diagram has a transition matrix $T^{(1)}$,

\begin{equation}
T^{(1)}=-U_{0}[2\pi ^{2}(\alpha ^{2}+|k_{i}-k_{f}^{\prime }|^{2})]
\end{equation}

$k_{i}$ and $k_{f}$ are the momentum of the initial and final states.
Similarly, the two step and the three step ladders will have the forms,

\begin{equation}
T^{(2)}=-2\pi ^{2}U_{0}^{2}\int \frac{d\stackrel{\_}{k}_{1}}{%
(k_{1}-k^{2}-i\epsilon )(\alpha ^{2}+|k_{i}-k_{1}^{\prime }|^{2})(\alpha
^{2}+|k_{1}-k_{f}^{\prime }|^{2})}
\end{equation}

\[
T^{(3)}=\int \int \frac{(U_{0}^{3}/4\pi ^{4})d\stackrel{\_}{k}_{1}d\stackrel{%
\_}{k}_{2}}{(k-k_{2}^{2}-i\epsilon )(\alpha ^{2}+|k_{2}-k_{1}^{\prime
}|^{2})(\alpha ^{2}+|k_{2}-k_{f}^{\prime }|^{2})(k-k_{1}^{2}-i\epsilon )}%
\cdot 
\]

\begin{equation}
\frac{1}{(k-k_{1}^{2}-i\epsilon )(\alpha ^{2}+|k_{i}-k_{1}^{\prime }|^{2})}
\end{equation}

The integrals in Eqs.(5,6) are carried out by the Dalitz integrals [12]. The
expansion coefficients $\{c_{i}\}$ in Eq.(1) are determined by fitting with
the experimental data to estimate the statistical weight factor of each
ladder diagram contributing the reaction. According to the usual formalism
for the dynamics of the particle reactions [13], the phase space integral is
an integration of the square of the transition matrix

$T(\{k_{i}\})$ over a set of allowed values $\{k_{i}\}$.

\begin{equation}
I_{n}(s)=\int \prod_{i}^{n}\frac{d^{3}k_{i}}{2E_{i}}\delta
^{4}(k_{a}+k_{b}-\sum_{i=1}^{n}k_{i})|T\{k_{i}\}|^{2}
\end{equation}

$k_{a},k_{b}$ are the momenta of the interacting particles in the initial
state, while $\{k_{i}\}$being the momenta of the particles in the final
state. For the problem under consideration of the elastic scattering, where
only two particles in the final state $n=2$, with the normalization
convention, the reaction cross section for $k_{a}+k_{b}\rightarrow
k_{1}+k_{2}$ is

\begin{equation}
\sigma (s)=\frac{1}{8\pi ^{2}\lambda ^{1/2}(s,m_{a}^{2},m_{b}^{2})}\int 
\frac{d^{3}k_{1}}{2E_{1}}\frac{d^{3}k_{2}}{2E_{2}}\delta
^{4}(k_{a}+k_{b}-k_{1}+k_{2})|T\{k_{1},k_{2}\}|^{2}
\end{equation}

$\lambda $ is a standard function defined as,$\lambda
(x,y,z)=(x-y-z)^{2}-4yz $ . One may immediately write the differential cross
section in the center of mass system CMS

\begin{equation}
\frac{d\sigma }{d\Omega }=\frac{1}{64\pi ^{2}s}\frac{k_{1}}{k_{a}}|T|^{2}
\end{equation}

It is more convenient to use the invariant cross section,

\begin{equation}
\frac{d\sigma }{dt}=\frac{1}{16\pi \lambda (s,m_{a}^{2},m_{b}^{2})}%
|T|^{2},\qquad t=-Q^{2}
\end{equation}

Eqs. (9),(10) as well as Eq.(1) are in a relevant form to compare with the
experimental data.

\section{The multi peripheral model (MPM)}

In this section, we shall deal with the problem of the particle production
in the (ep) inelastic scattering in view of the multi peripheral collision
[14]. A factorizable transition matrix $T$ is assumed in the form,

\begin{equation}
T=\prod_{i}^{n-1}T_{i}
\end{equation}

Each particle in the final state is produced at a specific peripheral
surface as shown by the Feynman diagram Fig.(2) with electromagnetic
transition matrix $T_{i}$, may be written in a suitable parametric form, 
\begin{equation}
T=\frac{1}{\alpha _{i}+t_{i}}
\end{equation}

where t$_{i}$ is the four vector momentum transfer square at the $i^{th}$
peripheral surface. $\alpha _{i}$ is the electromagnetic peripheral
parameter characterizing the surface number $i$, and determine to conserve
the total energy. The advantage of this technique is to reduce the many body
problem into $(n-1)$ iterative diagrams, each of them has only two particles
in the final state. For example, the $i^{th}$ diagram has two particles in
the final state, the first one is the particle number $(i+1)$, and the other
one has an effective mass $M_{i}$, equivalent to the rest of the i-particles
of the system. The square of the 4-vector momentum transfer $t_{i}$ is
kinematically calculated as,

\[
t_{i}=(p_{a}^{(i)}-p_{1}-....-p_{i})
\]

\begin{equation}
t_{i}=m_{i+1}^{2}+M_{i}^{2}-2E_{a}^{(i)}k_{i}^{0}-2P_{a}^{(i)}K_{i}\cos
\theta _{i}
\end{equation}

$m_{i+1}$ is the rest mass of the particle number $(i+1)$ produced in the $%
i^{th}$ iteration. $K_{i}$ and $k_{i}^{0}$ are the 3-vector momentum and the
total energy of the effective mass $M_{i}$. and are the corresponding
figures for the leading particle acting at the $i^{th}$ peripheral surface.
The recursion relation of $P_{a}^{(i)}$is given by,

\begin{equation}
P_{a}^{(i)}=\lambda ^{1/2}(M_{i}^{2},t_{i},m_{a}^{2})/2M_{i}
\end{equation}

The leading particle for the first peripheral surface is given by

\begin{equation}
P_{a}=P_{a}^{(n)}=\lambda ^{1/2}(s,m_{a}^{2},m_{b}^{2})/2\sqrt{s}
\end{equation}

The multi peripheral parameters $\{\alpha _{i}\}$ play important role in
converging the particles in phase space and consequently, control the energy
of the particles in final state. So that the values of $\{\alpha _{i}\}$ are
adjusted to conserve the total energy. The energy $E_{i}$ of the particle
number $i$ is related to its rapidity $y_{i}$ through the relation,

\[
E_{i}=m_{t}\cosh (y_{i})
\]

\begin{equation}
m_{t}=\sqrt{P_{t}^{2}+m_{i}^{2}}
\end{equation}

so that the total energy of the particles in the final state is

\begin{equation}
\zeta _{i}^{n}=\frac{1}{\sigma _{n}}\int m_{t}\cosh (y)(d\sigma /dy)dy
\end{equation}

$\zeta _{i}^{n}$ which is a function of the parameters $\{\alpha _{i}\}$
should be compared with the total center of mass energy$\sqrt{s}$ of the
initial state. We first start with $n=2$ to get $\alpha _{1}$, which is
inserted again in the case $n=3$ to get $\alpha _{2}$ and so on. These are
repeated sequentially to get the values of the rest parameters up to $\alpha
_{n-1}$. The values $\{\alpha _{i}\}$ depend on the particle multiplicity n
rather than the energy $\sqrt{s}$. The phase space integral $I_{n}(s)$ is
then calculated as in Eq.(7) after transforming the integral variables from $%
k_{i}$ to $t_{i}$,

\[
I_{n}(s)=\frac{(2\pi )^{n-1}}{2M_{n}}\prod_{i=3}^{n}\frac{1}{4P_{a}^{(i)}}%
\int_{\mu _{i}}^{M_{i}-m_{i}}dM_{i-1}\int_{t_{i-1}^{-}}^{t_{i-1}^{+}}\frac{1%
}{t_{i-1}+\alpha _{i-1})^{2}}dt_{i-1}\cdot 
\]

\begin{equation}
\frac{1}{4P_{a}^{(2)}}\int_{t_{1}^{-}}^{t_{1}^{+}}\frac{1}{t_{1}+\alpha
_{1})^{2}}dt_{1}
\end{equation}

\begin{equation}
\mu _{i}=\sum_{j=1}^{i}m_{j}
\end{equation}

$t_{i}^{\pm }$ are the upper and lower limits of $t_{i}$ corresponding to $%
cos(\theta _{i})=\pm 1.$

\begin{equation}
t_{_{i}}^{\pm }=m_{a}^{2}+m_{b}^{2}-2E_{a}E_{b}\pm 2P_{a}P_{b}
\end{equation}

Multiple integrals in Eq.(18) are carried out using a Monte Carlo program,
through which all possible distributions of the physical quantities are
easily found.

\section{Monte Carlo}

A Monte Carlo program GENE2 [15] is developed to simulate events according
to the multi peripheral diagrams. It includes 3-generators. The generator $%
Gn(s)$ for the multiplicity of particles in the final state, the generator $%
GM(s,n)$ for the invariant masses produced at the $n-1$ possible peripheral
surfaces and finally the dynamic generator $Gt(s,n,M)$ which generates the
values of the momentum transfer square $t_{i}$ of the $i^{th}$ surface
according to electromagnetic transition matrix. The program GENE2 executes
the 3- generators in ordered sequences, and simulates all the kinematical
variables of the n- particles in the final state.

A general form of the algorithm used to generate an event x with probability 
$f(x)$ is,

\begin{equation}
r=\int_{a}^{x}f(x)dx/\int_{a}^{b}f(x)dx
\end{equation}

Where $a$ and $b$ are the boundaries of the physical region at which x is
defined. $r$ is the default random number with uniform distribution between
the limits $\{0,1\}$.

\subsection{The multiplicity generator Gn(s)}

It is assumed that the multiplicity distribution $P(n)$ has a Gaussian form
[16],

\begin{equation}
P(n)=\frac{1}{\sqrt{2\pi }\sigma }\exp [-(n-\stackrel{-}{n})^{2}/\sigma ^{2}]
\end{equation}

where $\stackrel{-}{n}$ is the average multiplicity which depends on the
center of mass energy $\sqrt{\sigma }$ and $\sigma $ is the dispersion.

\begin{equation}
\stackrel{-}{n}=a\log (s)+b\qquad ,\sigma =c\,\stackrel{-}{n}+d
\end{equation}

the integration limits extend from zero to infinity. This makes Eq.(21) read;

\begin{equation}
r=Erf\,(n-\stackrel{-}{n})
\end{equation}

$Erf(x)$ is the error function. The solution $n$ of Eq. (24) defines the
multiplicity generator of the reaction.

\subsection{ The invariant mass generator GM(s,n)}

According to the multi peripheral diagrams Fig.(2), it is assumed that the
peripheral surface number i, produces the particle number $i+1$ and the
invariant mass $M_{i},$ equivalent to the effective mass of the system of
the rest of the $i-$particles. The value of $M_{i}$ satisfies the relation,

\begin{equation}
\sum m_{i}\leq M_{i}\leq M_{i+1}
\end{equation}

so that the algorithm of the generator GM(s,n) is,

\begin{equation}
M_{i}=\sum_{j}^{i}m_{j}+r(M_{i+1}-\sum_{j=1}^{i}m_{j})
\end{equation}

For a system of n-particle final state, we start with $M_{n}=\sqrt{s}$, then
generate the value of $M_{n-1}$. Eq.(26) is the recursion relation to
generate all values of $M_{i},\quad i=n-1,...,2$.

\subsection{The dynamic generator Gt(s,n,M)}

The algorithm used to generate the values of the square of the momentum
transfer $t_{i}$ should simulate a probability distribution which is
proportional to the square of the transition matrix $T_{i}$ defined by
Eq.(12). Inserting this in Eq.(21), we get the $t_{i}$ generator as,

\begin{equation}
t_{i}=\{r[(t_{i}^{+}+\alpha _{i})^{-1}-(t_{i}^{-}+\alpha
_{i})^{-1}+(t_{i}^{+}+\alpha _{i})^{-1}\}^{-1}-\alpha _{i}
\end{equation}

\section{Results and discussion}

\subsection{ The elastic scattering}

The problem of elastic scattering is treated in this article using the
ladder diagrams of multi steps. The transition matrix $T^{(n)}$ is
calculated for ladder diagrams for $n=1,2$ and $3$ using the Dalitz and
Feynman integrals. The results are demonstrated in Fig.(3). The potential
field acting at each vertex of the diagram is assumed to be of
electromagnetic nature with screening factor due to the pion current inside
the proton target. The screening effect limits the infinite range of the
coulomb potential. It is found that the terms of the transition matrix $%
T^{(n)}$ form a converging series. The first term of which decreases rapidly
with the momentum transfer square $t=-Q^{2}$. while the higher order terms
are slowly varying functions. The asymptotic behavior of $T^{(n)}$ follows a
power law at extremely high energy.

The expansion coefficients $\{c_{i}\}$ are determined by the fitting method
with the SLAC experimental data [4] in the range $t\backsim 1-3(GeV/c)^{2}$.
The results are shown in Fig.(4).

It is found that only two terms of the series are sufficient to represent
the reaction. The maximum probable diagram corresponds to $n=1$. On the
other hand, the SLAC-E-136 experiment, Fig.(5) in the momentum transfer
range $t\backsim 3-31(GeV/c)^{2}$ needs more terms of the series, with a
most probable diagram corresponds also to $n=1$. The values of the expansion
coefficients are given in Table (1).

\smallskip 

\begin{center}
Table(1) The branching ratios of multi photon exchanged in ep
collisions.\qquad \qquad \qquad 
\begin{tabular}{|c|c|c|c|}
\hline
Reaction & 1-Step & 2\_Steps & 3-Steps \\ \hline
Ref.[3] & 91 & 9 & - \\ \hline
SLAC-E-136 & 86 & 12 & 2 \\ \hline
\end{tabular}
\end{center}

\medskip

Figure (4) and (5) show good agreement between the experimental data and the
prediction of the multi photon exchange mechanism. It is clear that the
reactions of high momentum transfer needs more terms of ladder diagrams.

\subsection{The inelastic scattering.}

Here we use the iterative multi peripheral diagram as in Fig.(2-a,b). The
advantage of this is to simulate a complete inclusive reaction. We used the
Monte Carlo program GENE2 to generate the n-particles in the final state.
The multi peripheral parameters $\alpha _{i}$, $i=1,2,\cdot \cdot \cdot ,n-1$
are determined to conserve the total center of mass energy $\sqrt{s}$. The
values of the parameters $\{\alpha _{i}\}$ depend not only on $\sqrt{s}$ but
also on the degree of peripherality $i$. The first few peripheral surfaces
possess parameter values which confine the particle production in very
narrow cone. The production cone angle gets wider for higher order surfaces.
The values of $\alpha _{i}$ are displayed in Fig.(6) as a function of the
multiplicity $n$, of the number of particle in the final state for electron
lab energies $50,100$ and $200GeV$. In all cases the value of $\alpha _{i}$
decreases slowly with n, up to a critical multiplicity nc after which a
sudden drop is observed. A parametric relation is obtained for $\alpha _{i}$
as a function of the electron lab energy $E$ and $n$ as,

\[
\alpha _{n-1}=(0.226-0.015E+0.0003E^{2})n+0.441\log (E)-2.29,\quad n<n_{c}.
\]

\[
\alpha _{n-1}=(0.0019E-0.8685)\exp [(-.000688E+0.327276)n],\quad \;n>n_{c}
\]

\begin{equation}
n_{c}=0.4846E+1.0769
\end{equation}

The critical value $n_{c}$ corresponds to the peripheral surface at which
enough energy is transferred, that is sufficient to make phase transition
from the nuclear matter to the quark gluon state. The model is applied to
the data of the experiment SLAC-E-133 corresponding to electron lab energies 
$9.744,12.505,15.730,18.476$ and $20.999GeV.$ Fig.(7) shows the rapidity
distributions for particles produced at $9.744GeV$ electron lab energy as
calculated by (MPM) at peripheral surfaces $n=2,3,5$ and $10$. The result
for $n=2$ (two particle final state) shows two non symmetric clear peaks
corresponding to forward and backward emission. The electron peak is
relatively narrower than the proton one

. This case represents the most peripheral collision with low momentum
transfer. As the multiplicity increases the two peaks get broader and
interfere through each other. They completely interfere and form only one
peak at the case of high multiplicity $n=10$ corresponding to high momentum
transfer. This represents the most central collision. i.e. collision between
the electron and the proton core. The missing mass of the recoil nucleon is
also calculated through the Monte Carlo program GENE2, according to the
relation,

\begin{equation}
W^{2}=M_{p}^{2}+2M_{p}(E-E^{\prime })-4EE^{\prime }\sin ^{2}\frac{\theta }{2}
\end{equation}

\bigskip 

The differential cross section $d\sigma /d\Omega $ as a function of $W^{2}$
is compared in Fig.(8) with the experimental data of energies $9.744,12.505$
and $15.730GeV$. Fair agreement is obtained for the first reaction only. The
deviation increases as the electron energy increases. This may be due to the
fact that the model in hand has ignored the relative motion of the core
inside the nucleon target. This relative motion let the recoil nucleon not
has a unique value of missing mass, but instead, it posses a Gaussian -like
distribution around $W^{2}=M_{p}^{2}$. The peak is wider at higher $t$.

\section{Conclusion}

1- Data of elastic scattering may be reproduced in a wide range of the
energy transferred using the (MPEM) with an expandable transition matrix
representing ladder diagrams.

2- A transition matrix of order $n$, may be approximated as the power form
of the one step ladder diagram at extremely high energy.

3- The power series of the transition matrix contains number of terms
increasing with the energy transferred of the reaction.

4- Inelastic ep scattering are successfully described by the multi
peripheral model.

5- The peripheral parameters are the dynamic parameters, which control the
convergence of the particles in phase space. A critical value of peripheral
parameter determines the critical surface at which phase transition may
occur from the nuclear matter density to the quark gluon phase.

6- The rapidity distribution of the particles in the final state shows two
peaks representing the forward and backward production. The peaks interfere
together as the multiplicity increases in the final state.

\end{document}